# High-Throughput Precision Phenotyping of Left Ventricular Hypertrophy with Cardiovascular Deep Learning

Grant Duffy[1,#], Paul P Cheng[2,#], Neal Yuan[1,#], Bryan He[3], Alan C. Kwan[1], Matthew J. Shun-Shin[4], Kevin M. Alexander[2], Joseph Ebinger[1], Matthew P. Lungren[5], Florian Rader[1], David H. Liang[2], Ingela Schnittger[2], Euan A. Ashley[2], James Y. Zou[3,6], Jignesh Patel[1], Ronald Witteles[2], Susan Cheng[1,*], David Ouyang[1,7,*]

1. Department of Cardiology, Smidt Heart Institute, Cedars-Sinai Medical Center
2. Department of Medicine, Division of Cardiology, Stanford University
3. Department of Computer Science, Stanford University
4. National Heart and Lung Institute, Imperial College London
5. Department of Radiology, Stanford University
6. Department of Biomedical Data Science, Stanford University
7. Division of Artificial Intelligence in Medicine, Cedars-Sinai Medical Center

\# Co-first author
\* Co-senior author

Correspondence: Susan.cheng@cshs.org and David.ouyang@cshs.org

Word Count: 2,846

**Abstract**

Left ventricular hypertrophy (LVH) results from chronic remodeling caused by a broad range of systemic and cardiovascular disease including hypertension, aortic stenosis, hypertrophic cardiomyopathy, and cardiac amyloidosis. Early detection and characterization of LVH can significantly impact patient care but is limited by under-recognition of hypertrophy, measurement error and variability, and difficulty differentiating etiologies of LVH. To overcome this challenge, we present EchoNet-LVH – a deep learning workflow that automatically quantifies ventricular hypertrophy with precision equal to human experts and predicts etiology of LVH. Trained on 24,804 echocardiogram videos, our model accurately measures intraventricular wall thickness (mean absolute error [MAE] 1.4mm, 95% CI 1.2-1.5mm), left ventricular diameter (MAE 2.4mm, 95% CI 2.2-2.6mm), and posterior wall thickness (MAE 1.2mm, 95% CI 1.1-1.3mm) and classifies cardiac amyloidosis (area under the curve of 0.83) and hypertrophic cardiomyopathy (AUC 0.98) from other etiologies of LVH. In external datasets from independent domestic and international healthcare systems, EchoNet-LVH accurately quantified ventricular parameters ($R^2$ of 0.96 and 0.90 respectively) and detected cardiac amyloidosis (AUC 0.79) and hypertrophic cardiomyopathy (AUC 0.89) on the domestic external validation site. Leveraging measurements across multiple heart beats, our model can more accurately identify subtle changes in LV geometry and its causal etiologies. Compared to human experts, EchoNet-LVH is fully automated, allowing for reproducible, precise measurements, and lays the foundation for precision diagnosis of cardiac hypertrophy. As a resource to promote further innovation, we also make publicly available a large dataset of 12,000 annotated echocardiogram videos.

**Background**

Despite rapidly advancing developments in targeted therapeutics and genetic sequencing[1,2], persistent limits in the accuracy and throughput of clinical phenotyping has led to a widening gap between the potential and the actual benefits realized by precision medicine. This conundrum is exemplified by current approaches to assessing morphologic alterations of the heart[3,4]. If reliably identified, certain cardiac diseases (e.g. cardiac amyloidosis and hypertrophic cardiomyopathy) could avoid misdiagnosis and receive efficient treatment initiation with specific targeted therapies. The ability to reliably distinguish between cardiac disease types of similar morphology but different etiology would also enhance specificity for linking genetic risk variants and determining mechanisms.

The heart is a dynamic organ capable of remodeling and adapting to physiologic stress and extra-cardiac perturbation. Both intrinsic cardiac disease as well as systemic insults can result in similar presentations of left ventricular hypertrophy (LVH) that are difficult to distinguish on routine imaging by human observation. Pressure overload from long standing hypertension and aortic stenosis can cause cardiac remodeling to compensate for additional physiologic work, while hypertrophic cardiomyopathy and cardiac amyloidosis can similarly manifest with an increase in left ventricular mass but in the absence of physiologic stress.

In addition to the presence of LVH, the degree of ventricular thickness also has significant prognostic value in many diseases[5–7]. Ventricular wall thickness is used to risk stratify patients for risk of sudden cardiac death and help determine which patients should undergo defibrillator implantation[7]. Nevertheless, quantification of ventricular thickness remains subject to significant intra- and inter-provider variability across imaging modalities[8,9]. Even with the high image resolution and signal-to-noise ratio of cardiac magnetic resonance imaging, there is significant test-retest variability due to the laborious, manual nature of wall thickness measurement[10,11]. Although abundant[12,13], low cost, and without ionizing radiation, echocardiography relies on expert interpretation and its accuracy is dependent on careful application of measurement techinques[14,15].

Recent work has shown that deep learning applied to medical imaging can identify clinical phenotypes beyond conventional image interpretation and with higher accuracy than human experts[16–19]. As the most common form of cardiovascular imaging and the basis of society guidelines for diagnosing hypertrophy[7], we hypothesize that echocardiography, when enhanced with artificial intelligence models, can provide additional value in understanding disease states by predicting both the presence of left ventricular hypertrophy in a screening population as well as the potential etiology of diagnosis. To overcome current limitations in the assessment of ventricular hypertrophy and disease diagnosis, we present EchoNet-LVH, an end-to-end deep learning approach for labelling the left ventricle dimensions, quantifying ventricular wall thickness, and predicting etiology of LVH. We first conducted frame-level semantic segmentation of the left ventricular wall thickness from parasternal long axis echocardiogram videos and then performed beat-to-beat evaluation of ventricular hypertrophy. After identifying left ventricular hypertrophy, we used a three-dimensional convolutional neural network with residual connections to predict the etiology of the LVH, including predictions for cardiac amyloidosis and aortic stenosis among a background of other hypertrophic diseases.

## Results

EchoNet-LVH has two components (Fig. 1). First, we designed a deep learning model with atrous convolutions for semantic segmentation of parasternal long axis (PLAX) echocardiogram videos and identification of the intraventricular septum (IVS), left ventricular internal dimension (LVID), and left ventricular posterior wall (LVPW). With atrous convolutions to capture longer range features, full resolution PLAX frames were used as input images for higher resolution assessment of LVH. Given the tedious nature of annotation, the standard clinical workflow often only labels one or two frames of a video, however we generalize these sparse annotations into measurement predictions for every frame of the entire video to allow for beat-to-beat estimation of ventricular wall thickness and dimensions.

After detection of LVH, identifying the specific etiology (e.g. infiltrative disease, inherited cardiomyopathies, or chronic elevated afterload) can help guide therapy. We trained a video-based CNN model with spatiotemporal convolutions to predict etiology of LVH. Integrating spatial as well as temporal information, our model expands upon our prior work with video-based model interpretation of echocardiograms and classifies videos based on probability of hypertension, aortic stenosis, hypertrophic cardiomyopathy, or cardiac amyloidosis as etiology of ventricular hypertrophy. We additionally performed a video-based model architecture and hyperparameter search to identify the optimal base architecture for EchoNet-LVH (Extended Data Figure 1). EchoNet-LVH was trained on a dataset of 24,804 echocardiogram videos from Stanford Health Care (SHC), and then evaluated on held out test cohorts from SHC, Cedars-Sinai Medical Center (CSMC), and Unity Imaging Collaborative.

**Evaluation of Hypertrophy Detection**
From the held-out test dataset from SHC (n = 1,200) not seen during model training, EchoNet-LVH predicted ventricular dimensions with a $R^2$ of 0.97 compared to annotations by human experts (Fig. 2). EchoNet-LVH had a mean absolute error (MAE) of 1.2mm for IVS, 2.4mm for LVID, and 1.4mm for LVPW. This compares favorably with clinical inter-provider variation, which had a MAE of 1.3mm for IVS, 3.7mm for LVID, and 1.3mm for LVPW. To assess the cross-healthcare-system and international reliability of the model, EchoNet-LVH was additionally tested, without any tuning, on an external test dataset of 1,791 videos from Unity Imaging Collaborative and 3,660 videos from CSMC. On the Unity external test dataset, EchoNet-LVH showed a robust prediction accuracy with an overall $R^2$ of 0.90, MAE of 1.6mm for IVS, 3.6mm for LVID, and 2.1 mm for LVPW.

A rapid, high-throughput automated approach allows for measurement of every individual frame that would be tedious for manual tracing (Fig 2). Differences in filling time and irregularity in the heart rate can cause variation in measurement but beat-to-beat model assessment can provide higher fidelity overall assessments. While the SHC and Unity datasets were directly compared on sparsely annotated individual frames, we evaluated EchoNet-LVH's beat-to-beat evaluation on the CSMC dataset in comparison with study-level annotations of ventricular dimensions by expert clinicians. In this dataset, human measurements were not associated with specific frames of the echocardiogram video, and beat-to-beat analysis had to predict diastole in each heart beat and average across the entire video. On the CSMC external test dataset, EchoNet-LVH showed a robust prediction accuracy with an overall $R^2$ of 0.96, MAE of 1.7mm (95% CI 1.6–1.8mm) for IVS, 3.8mm (95% CI 3.5-4.0mm) for LVID, and 1.8mm (95% CI 1.7-2.0mm) for LVPW with beat-to-beat evaluation.

**Prediction of Etiology of Hypertrophy**

The etiology derivation, validation, and test cohorts from SHC had 6,215, 787, and 765 videos respectively. On the held-out test cohort, EchoNet-LVH classifies cardiac amyloidosis with an AUC of 0.83, hypertrophic cardiomyopathy with an AUC 0.98, and aortic stenosis with an AUC 0.89 from other etiologies of LVH. On an external test dataset of 2,351 A4c videos from CSMC with 358 videos of cardiac amyloidosis, 146 videos of aortic stenosis, 468 videos of hypertrophic cardiomyopathy, and 1,379 videos of other etiologies of LVH, EchoNet-LVH had an AUC of 0.79 for predicting cardiac amyloidosis and an AUC of 0.89 for hypertrophic cardiomyopathy.

**Discussion**

EchoNet-LVH is a deep learning system for quantifying left ventricular hypertrophy on echocardiography with automated prediction of LVH etiology. EchoNet-LVH achieves state-of-the-art performance in assessment of ventricular thickness and diameter, within the variance in clinical test-retest assessment, and aids in detection of subtle phenotypes that are particularly challenging for human readers. Integrating the steps of identifying ventricular hypertrophy and predicting its etiology, EchoNet-LVH provides a fully automated workflow for disease screening via the most frequently used form of cardiac imaging. Such assimilation of diagnostic processes with use of the widest available clinical imaging modality may substantially improve opportunities for more targeted and effective cardiovascular care.

Whether used to triage patients for systemic therapy for cardiac amyloidosis, targeted medications for hypertrophic cardiomyopathy, or percutaneous valve replacement for aortic stenosis, EchoNet-LVH can facilitate early expedited care for heart disease with ventricular remodeling. Validated in both domestic and international datasets not seen during model training, EchoNet-LVH's performance in assessing ventricular thickness was robust across continents, clinical practice patterns, and instrumentation for image acquisition. Additionally, EchoNet-LVH performs these tasks in real time with only one GPU; each prediction takes only 24.6ms per frame and is more rapid than human assessment, allowing for the possibility of real-time screening of cardiovascular disease in a clinical setting.

Over the last two decades, the cost of genetic sequencing has dropped precipitously[2], making possible the application of genetics clinical practice and polygenic risk scores for complex diseases[20]. While sequencing technology has followed the trajectory of Moore's law in cost-effectiveness and efficiency[1], clinical phenotyping has only risen in price and been stagnant in throughput[21]. As such, phenotyping has become the bottleneck in future applications of genomics to clinical medicine, and the heterogeneity and imprecision of clinical measurements limit the power of genetic studies. Given the labor cost of clinical phenotyping and intra-clinician variability in annotation, previous genetic studies from echocardiography parameters have shown only small genetic associations, some of which have not been replicable[3,4,22–24]. The application of precise, high-throughput measurements through deep learning of readily available imaging such as echocardiograms could open avenues for further exploration only available through the combination of low-cost phenotyping and low-cost sequencing.

Our model was trained on videos obtained by trained sonographers at an academic medical center that reflect the variation in clinical practice. With expansion in the use of point-of-care ultrasound for evaluation of cardiac function by non-cardiologists, further work needs to be done to understand model performance with input videos of more-variable quality and acquisition expertise as well as comparison with other imaging modalities. Our analyses across health systems suggest that EchoNet-LVH is robust to variation in practice patterns across continents; however, prospective deployment and testing of AI systems in diverse clinical environments remain to be done.

Expanding our prior work, we collaborated with stakeholders across Stanford Medicine to release our training dataset of 12,000 de-identified PLAX echocardiogram videos as a resource for the medical machine learning community for future comparison and validation of deep learning models. This expands our prior dataset of 10,030 videos to a total of 22,030 echocardiogram videos made publicly available, the largest dataset release of labelled medical videos with matched human expert annotations. We hope this dataset will facilitate new echocardiogram and medical video-based machine learning approaches. We have also released the full code for our algorithm and data-processing workflow.

Our results represent an important step towards the automated assessment of cardiac structures in echocardiogram videos through deep learning. Combined with prior work assessing cardiac function, EchoNet-LVH augments current methods for assessing cardiac form and structure to provide more holistic evaluation of cardiovascular disease. With improved precision to detect ventricular remodeling and cardiac dysfunction, artificial intelligence systems have the potential enable earlier detection and treatment of subclinical cardiovascular disease.

## Methods

**Data Curation**

A standard full resting echocardiogram study consists of a series of 50-100 videos and still images visualizing the heart from different angles, locations, and image acquisition techniques (2D images, tissue Doppler images, color Doppler images, and others). In this study, relevant parasternal long axis (PLAX) and apical-4-chamber (A4C) 2D videos were extracted from each study. Human expert annotations of IVS, LVID, and LVPW were used as training labels to assess ventricular hypertrophy. From SHC, PLAX videos were split 9,600, 1,200, and 1,200 patients respectively for the training, validation, and test sets. An additional 7,767 studies of SHC patients with defined disease characteristics, including cardiac amyloidosis, hypertrophic cardiomyopathy, severe aortic stenosis, hypertension and other phenotypes of LVH were identified through the electronic healthcare system. From these studies, the A4C videos were extracted and used as input data for the hypertrophic disease classification task. Videos were processed in a previously described automated preprocessing workflow removing identifying information and human labels[16]. Videos were spot checked for quality control, confirm view classification, and exclude videos with color Doppler. This research was approved by the Stanford University and Cedars-Sinai Medical Center Institutional Review Boards.

**Domestic and International External Health Care System Test Datasets**

Transthoracic echocardiogram studies from CSMC and the Unity Imaging Collaborative[11] were used to evaluate EchoNet-LVH's performance in identifying key points in PLAX videos and measuring ventricular dimensions. Previously described methods were used to identify PLAX and apical-4-chamber view videos and convert DICOM files to AVI files[22]. In total, we extracted 3,660 total videos from CSMC as a domestic held out test dataset. Labeled echocardiogram images from the Unity Imaging Collaborative were used as a separate held-out international test dataset not seen during model training. Given that EchoNet-LVH was trained on a separate training set, the entirety of the Unity Imaging dataset with PLAX annotations were used as an international external test set. These echocardiogram videos were obtained from 7 British echocardiography labs and were already pre-processed with annotated frames saved as PNG files.

**EchoNet-LVH Development and Training**

Model design and training was done in Python using the PyTorch deep learning library. A modified DeepLabV3[25] architecture trained on parasternal long axis images to minimize a weighted mean square error loss was used to identify key points used for measuring ventricular dimensions. An Adam optimizer was used with a learning rate of 0.001 was used and the model was trained for 50 epochs with early stopping based on the validation loss. For video-based disease classification, an 18-layer ResNet3D[26] architecture was used to classify videos as either amyloid or not amyloid. This model was trained to minimize binary cross-entropy loss using an Adam optimizer with a learning rate of 0.01. The model was trained for 100 epochs with a batch size of 14 with early stopping based on AUC on the validation set. We evaluated different video lengths, resolutions, and temporal resolutions as hyperparameters to optimize model performance (Extended Figure 1). Computational cost was evaluated using one NVIDIA GeForce GTX 3090.

**Test Time Augmentation with Beat-by-Beat Assessment**

For PLAX measurement prediction, test-time augmentation was performed with aggregating predictions across the entire echocardiogram video. The predicted LVID measurement was used to determine frames of peak systole and peak diastole. These frames were used to generate systolic and diastolic measurements for every beat of an echocardiogram video. These measurements were compared to the human labelled measurements (Fig. 2). Variation from beat to beat in a single echocardiogram is used to evaluate the precision of the method.

**Comparison with Human Measurement**

Using the reporting database of Stanford Echocardiography Laboratory, we identified paired studies of the same patient for which the reviewing cardiologist determined there was no significant change from the current study to the prior study. Of these studies with clinical stability, we analyzed the subset of 23,874 studies for which LVID, IVS, and LVPW at diastole was measured for both the current and subsequent study. The variance in measurement between the previous and subsequent study was used as a surrogate of clinician expert variation and compared with EchoNet-LVH variation.

## Statistical Analysis

Confidence intervals were computed using 10,000 bootstrapped samples and obtaining 95 percentile ranges for each prediction. The performance of the semantic segmentation task was evaluated comparing the length of LVID, LVPW, and IVS to human labels in the hold-out test dataset. The centroid of each predicted key point was used to calculate measurements.

## Acknowledgements

D.O. is supported by the National Institutes of Health grants K99-HL157421, N.Y. is supported by 5T32HL116273-07.

## Author Contributions

G.D, P.P.C, N.Y., I.J., J.E., D.O. retrieved, preprocessed, and quality-controlled videos and merged electronic medical record data. G.D., B.H., J.Y.Z., D.O., developed and trained the deep learning algorithms, performed statistical tests, and created all the figures. M.P.L., D.H.L., I.S., D.O. coordinated public release of the deidentified echocardiogram dataset. G.D., P.P.C, N.Y., S.C., D.O wrote the manuscript with feedback from all authors.

## Competing Interests

The authors declare no competing financial interests.

## Data Availability

This project introduces the EchoNet-LVH Dataset, a publicly available dataset of de-identified echocardiogram videos available at: https://echonet.github.io/plax/

## Code Availability

The code for EchoNet-LVH is available at: https://github.com/echonet/lvh

# References

1. Mardis, E. R. A decade's perspective on DNA sequencing technology. *Nature* **470**, 198–203 (2011).

2. DNA Sequencing Costs: Data. https://www.genome.gov/about-genomics/fact-sheets/DNA-Sequencing-Costs-Data.

3. Lindpaintner, K. *et al.* Absence of association or genetic linkage between the angiotensin-converting-enzyme gene and left ventricular mass. *N. Engl. J. Med.* **334**, 1023–1028 (1996).

4. Vasan, R. S. *et al.* Genetic variants associated with cardiac structure and function: a meta-analysis and replication of genome-wide association data. *JAMA* **302**, 168–178 (2009).

5. Lee, G. Y. *et al.* Cardiac amyloidosis without increased left ventricular wall thickness. *Mayo Clin. Proc.* **89**, 781–789 (2014).

6. Pagourelias, E. D. *et al.* Echo Parameters for Differential Diagnosis in Cardiac Amyloidosis. *Circ. Cardiovasc. Imaging* **10**, e005588 (2017).

7. Authors/Task Force members *et al.* 2014 ESC Guidelines on diagnosis and management of hypertrophic cardiomyopathy: the Task Force for the Diagnosis and Management of Hypertrophic Cardiomyopathy of the European Society of Cardiology (ESC). *Eur. Heart J.* **35**, 2733–2779 (2014).

8. Phelan, D. *et al.* Comparison of Ventricular Septal Measurements in Hypertrophic Cardiomyopathy Patients Who Underwent Surgical Myectomy Using Multimodality Imaging and Implications for Diagnosis and Management. *Am. J. Cardiol.* **119**, 1656–1662 (2017).

9. Angeli, F. *et al.* Day-to-day variability of electrocardiographic diagnosis of left ventricular hypertrophy in hypertensive patients. Influence of electrode placement. *J. Cardiovasc. Med.* **7**, 812–816 (2006).

10. Augusto, J. B. *et al.* Diagnosis and risk stratification in hypertrophic cardiomyopathy using machine learning wall thickness measurement: a comparison with human test-retest performance. *Lancet Digit Health* **3**, e20–e28 (2021).

11. Howard, J. P. *et al.* Automated Left Ventricular Dimension Assessment Using Artificial Intelligence Developed and Validated by a UK-Wide Collaborative. *Circ. Cardiovasc. Imaging* **14**, e011951 (2021).

12. American College of Cardiology Foundation Appropriate Use Criteria Task Force *et al.* ACCF/ASE/AHA/ASNC/HFSA/HRS/SCAI/SCCM/SCCT/SCMR 2011 Appropriate Use Criteria for Echocardiography. A Report of the American College of Cardiology Foundation Appropriate Use Criteria Task Force, American Society of Echocardiography, American Heart Association, American Society of Nuclear Cardiology, Heart Failure Society of America, Heart Rhythm Society, Society for Cardiovascular Angiography and Interventions, Society of Critical Care Medicine, Society of Cardiovascular Computed Tomography, Society for Cardiovascular Magnetic Resonance American College of Chest Physicians. *J. Am. Soc. Echocardiogr.* **24**, 229–267 (2011).

13. Popescu, B. A. *et al.* Updated standards and processes for accreditation of echocardiographic laboratories from The European Association of Cardiovascular Imaging: an executive summary. *Eur. Heart J. Cardiovasc. Imaging* **15**, 1188–1193 (2014).

14. Leibundgut, G. *et al.* Dynamic assessment of right ventricular volumes and function by real-time three-dimensional echocardiography: a comparison study with magnetic resonance imaging in 100 adult patients. *J. Am. Soc. Echocardiogr.* **23**, 116–126 (2010).

15. Farsalinos, K. E. *et al.* Head-to-Head Comparison of Global Longitudinal Strain Measurements among Nine Different Vendors: The EACVI/ASE Inter-Vendor Comparison Study. *J. Am. Soc. Echocardiogr.* **28**, 1171–1181, e2 (2015).

16. Ouyang, D. *et al.* Video-based AI for beat-to-beat assessment of cardiac function. *Nature* **580**, 252–256 (2020).

17. Poplin, R. *et al.* Prediction of cardiovascular risk factors from retinal fundus photographs via deep learning. *Nat Biomed Eng* **2**, 158–164 (2018).

18. Ghorbani, A. *et al.* Deep learning interpretation of echocardiograms. *NPJ Digit Med* **3**, 10 (2020).

19. Raghunath, S. *et al.* Prediction of mortality from 12-lead electrocardiogram voltage data using a deep neural network. *Nat. Med.* (2020) doi:10.1038/s41591-020-0870-z.

20. Eisenstein, M. Ranking the risk of heart disease. *Nature* **594**, S6–S7 (2021).

21. Martin, A. B., Hartman, M., Lassman, D., Catlin, A. & National Health Expenditure Accounts Team. National Health Care Spending In 2019: Steady Growth For The Fourth Consecutive Year. *Health Aff.* **40**, 14–24 (2021).

22. Wild, P. S. *et al.* Large-scale genome-wide analysis identifies genetic variants associated with cardiac structure and function. *J. Clin. Invest.* **127**, 1798–1812 (2017).

23. Kupari, M. *et al.* Associations between human aldosterone synthase (CYP11B2) gene polymorphisms and left ventricular size, mass, and function. *Circulation* **97**, 569–575 (1998).

24. Gomez-Angelats, E. *et al.* Lack of association between ACE gene polymorphism and left ventricular hypertrophy in essential hypertension. *J. Hum. Hypertens.* **14**, 47–49 (2000).

25. Yurtkulu, S. C., Şahin, Y. H. & Unal, G. Semantic Segmentation with Extended DeepLabv3 Architecture. in *2019 27th Signal Processing and Communications Applications Conference (SIU)* 1–4 (2019).

26. Tran, D. *et al.* A Closer Look at Spatiotemporal Convolutions for Action Recognition. *arXiv [cs.CV]* (2017).

# Figures

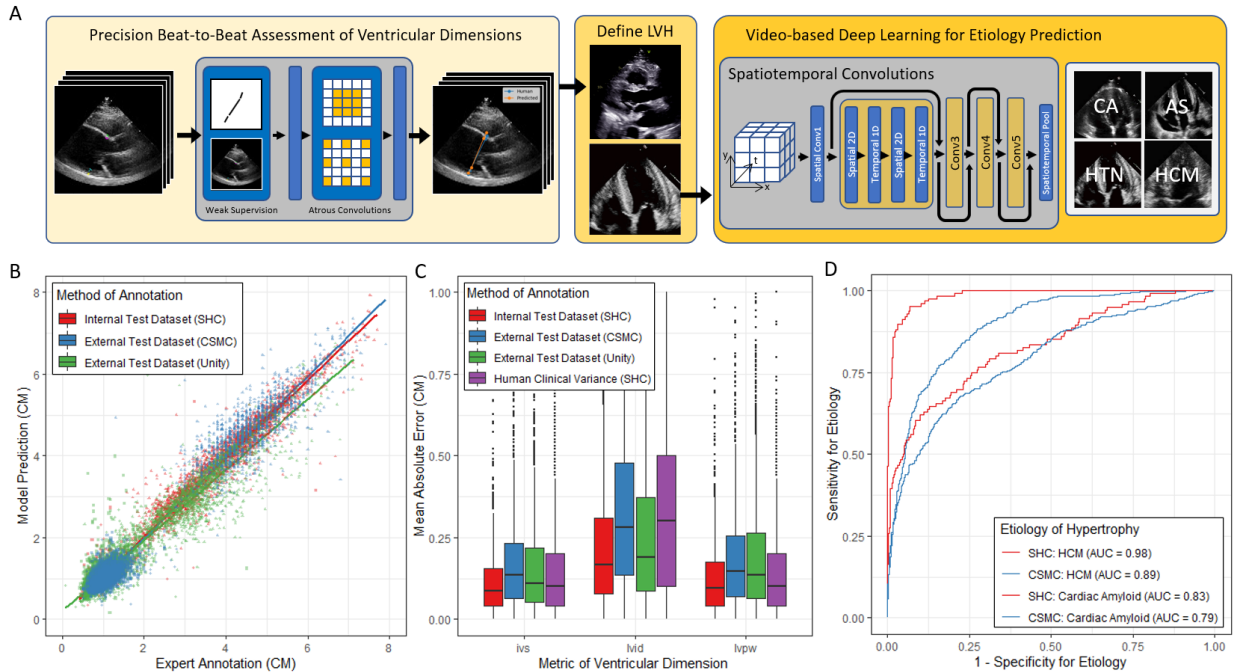

**Figure 1. EchoNet-LVH workflow**. A. For each patient, EchoNet-LVH uses parasternal long axis echocardiogram video as input to derive key points and precisely estimate ventricular dimensions. After identifying patients with LVH, EchoNet-LVH used a video-based architecture to distinguish between common causes of LVH. B. Correlation of human annotations vs. model predictions for ventricular dimensions in datasets from three healthcare systems (n = 1,200 for SHC, n = 1,309 for CSMC , and n = 1,791 for Unity). C. Model variation on datasets from three healthcare systems compared to human clinical annotation variation. Boxplot represents the mean as a thick line, 25th and 75th percentiles as upper and lower bounds of the box, and individual points for instances greater than 1.5 times the interquartile range from the mean. D. Receiver operating characteristic curves for diagnosis of amyloidosis on Stanford validation (n = 813) and test (n = 812) datasets. HCM = Hypertrophic cardiomyopathy, SHC = Stanford Health Care, CSMC = Cedars-Sinai Medical Center.

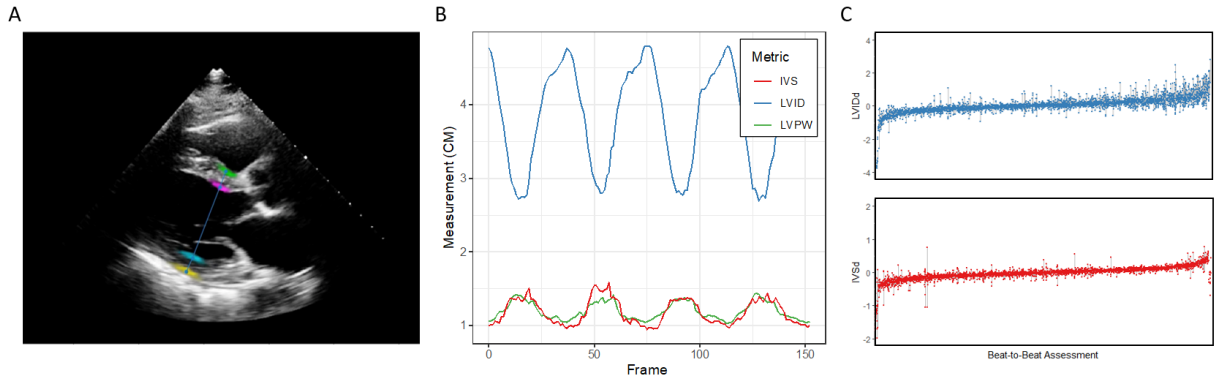

**Figure 2. Beat-to-beat evaluation of ventricular dimensions.** A. Model prediction of key points (in green, purple, blue, and yellow) on individual frame of parasternal long video. B. Frame-by-frame prediction of wall thickness and ventricular dimension and automated detection of systole and diastole allowing for beat-to-beat prediction of ventricular hypertrophy. C. Waterfall plot of individual video variation in beat-to-beat evaluation of ventricular hypertrophy (n = 2,320) across the internal test dataset. Each video is represented by multiple points along a line representing the measurement of each beat and a line signifying the range of predictions.

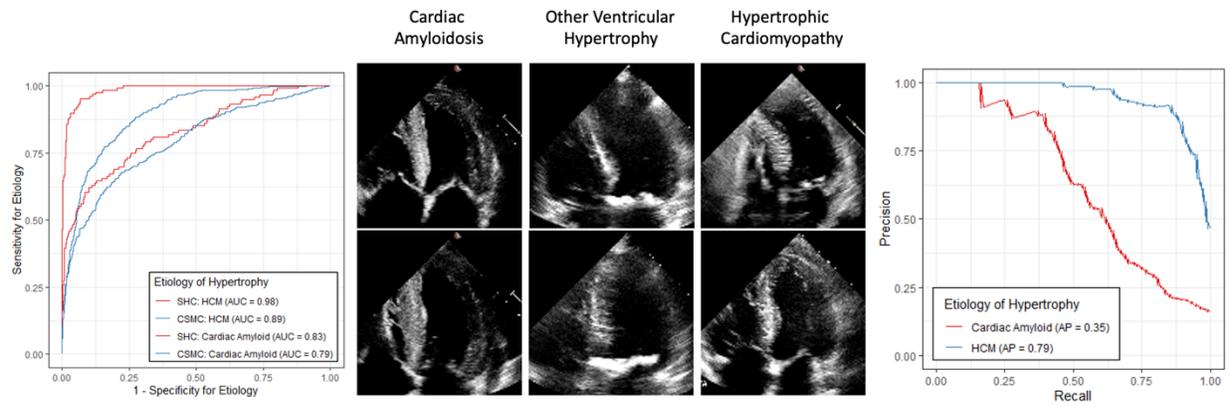

**Figure 3. Performance of Disease Etiology Classification.** A. Receiver operating characteristic curves for detection of cardiac amyloidosis and aortic stenosis in SHC internal test dataset (n = 765) and CSMC external test set (n = 2351). B. Representative images for selected cases and controls for each etiology. C. Precision-recall curves for detection of amyloidosis and hypertrophic cardiomyopathy in SHC test dataset (n = 765).

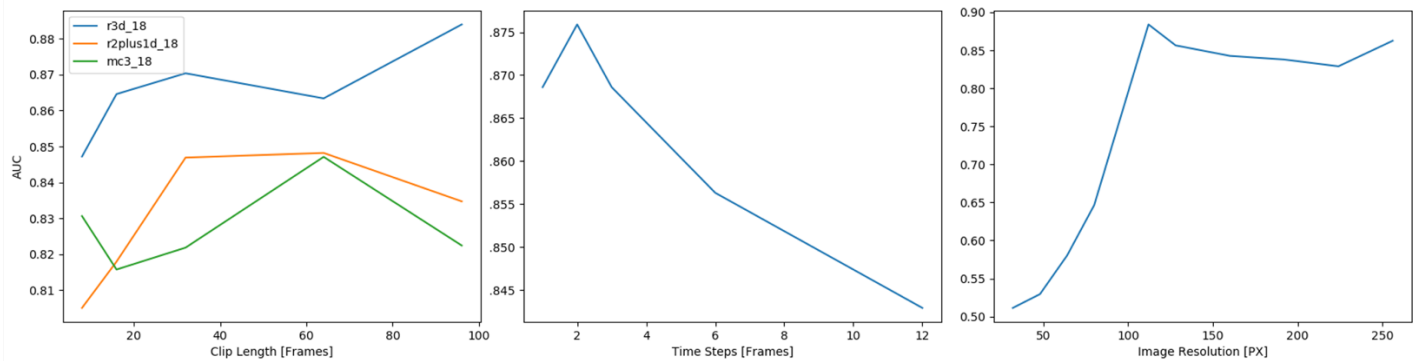

**Extended Figure 1. Hyperparameter Search of Video-based Classification Model** Hyperparameter search for model architecture (R3D, which is used by EchoNet-LVH for hypertrophy etiology classification, R2+1D, and MC3), input video clip length (8, 16, 32, 64, 96 frames) and impact on model processing time and model memory usage.

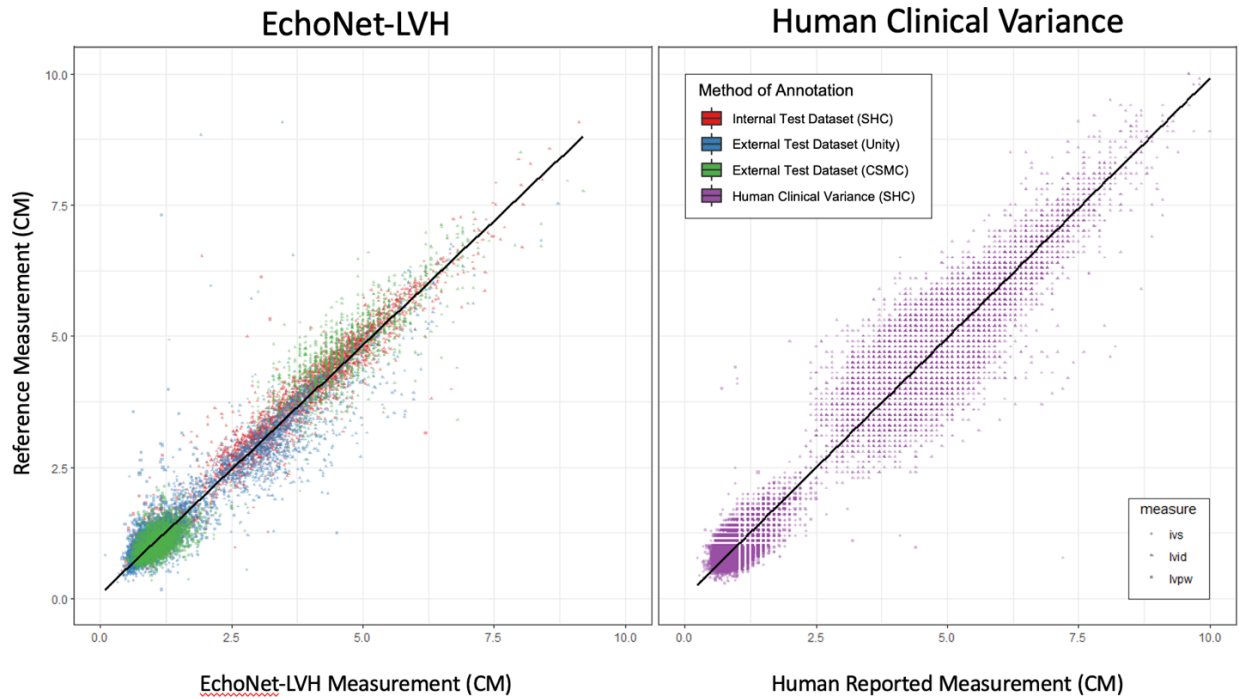

**Extended Figure 2. Comparison of Model Performance with Human Variation** A. Correlation of EchoNet-LVH measurements vs. human annotations for ventricular dimensions in datasets from three healthcare systems (n = 1,200 for SHC, n = 1,309 for CSMC, and n = 1,791 for Unity). B. Correlation of clinician reported measurements compared to prior study for studies without significant change (n = 23,874 at SHC).

|  | Apical 4 Chamber | | | | PLAX |
| --- | --- | --- | --- | --- | --- |
|  | Amyloid | HCM | AS | LVH |  |
| n | 358 | 468 | 146 | 1379 | 1309 |
| Age (mean (SD)) | 74.21 (9.26) | 55.10 (14.31) | 76.60 (8.65) | 72.57 (13.27) | 62.80 (17.16) |
| Male (%) | 320 (89.4) | 221 (47.2) | 107 (73.3) | 972 (70.5) | 808 (61.7) |
| Race/Ethnicity, n (%) |  |  |  |  |  |
|   American Indian | 0 (0.0) | 0 (0.0) | 0 (0.0) | 4 (0.3) | 4 (0.3) |
|   Asian | 7 (2.0) | 19 (4.1) | 12 (8.2) | 67 (4.9) | 85 (6.5) |
|   Black | 74 (20.7) | 21 (4.5) | 11 (7.5) | 247 (17.9) | 239 (18.3) |
|   Hispanic | 12 (3.4) | 111 (23.7) | 16 (11.0) | 127 (9.2) | 178 (13.6) |
|   Non-Hispanic White | 259 (72.3) | 290 (62.0) | 92 (63.0) | 827 (60.0) | 697 (53.2) |
|   Other | 6 (1.7) | 27 (5.8) | 15 (10.3) | 98 (7.1) | 83 (6.3) |
| Atrial Fibrillation, n (%) | 203 (56.7) | 68 (14.5) | 14 (9.6) | 345 (25.0) | 315 (24.1) |
| Congestive Heart Failure, n (%) | 307 (85.8) | 81 (17.3) | 19 (13.0) | 469 (34.0) | 451 (34.5) |
| Hypertension, n (%) | 195 (54.5) | 161 (34.4) | 61 (41.8) | 873 (63.3) | 543 (41.5) |
| Diabetes Mellitus, n (%) | 49 (13.7) | 19 (4.1) | 27 (18.5) | 383 (27.8) | 248 (18.9) |
| Coronary Artery Disease, n (%) | 174 (48.6) | 78 (16.7) | 31 (21.2) | 557 (40.4) | 369 (28.2) |
| Chronic Kidney Disease, n (%) | 130 (36.3) | 30 (6.4) | 18 (12.3) | 335 (24.3) | 257 (19.6) |
| Ejection Fraction (mean (SD)) | 51.96 (17.17) | 67.21 (9.97) | 60.99 (11.18) | 58.44 (13.25) | 55.92 (15.67) |
| LVPWd (mean (SD)) | 1.52 (0.52) | 1.16 (0.22) | 1.44 (0.13) | 1.47 (0.17) | 1.09 (0.25) |
| IVSd (mean (SD)) | 1.60 (0.52) | 1.57 (0.58) | 1.49 (0.28) | 1.51 (0.28) | 1.12 (0.29) |
| LVIDd (mean (SD)) | 4.32 (0.72) | 4.20 (0.67) | 4.84 (0.94) | 4.58 (0.88) | 4.70 (0.90) |
| LVIDs (mean (SD)) | 3.22 (0.87) | 2.65 (0.63) | 3.20 (0.88) | 3.11 (0.80) | 3.29 (1.05) |

**Extended Data Table 1. Baseline Characteristics of Patients with Cedars-Sinai Medical Center echocardiograms used for external validation.** PLAX = Parasternal Long Axis view videos. HCM = Hypertrophic cardiomyopathy. AS = Aortic stenosis. LVH = left ventricular hypertrophy.

## Supplemental Methods

**Ventricular Dimension Calculation**

Sparse human expert annotations were used for training of semantic segmentation task of identifying key points along ventricular septum and posterior wall to calculate IVS, LVID, and LVPW. For each video, only one frame was labeled at diastole and systole. Each key point (four in total) was mapped to four output channels of a DeepLabV3 architecture with a gaussian sampling to simulate human variation. The gaussian sampling was performed by the data loader during training and standard deviation of the gaussian sampling was empirically sampled as a model hyperparameter. The labels were four channel images with all zeros except ones for pixels nearest to the ground truth measurement point in each channel.

The loss function used to train the PLAX model is a modified mean squared error loss augmented with the L2 loss for point location error and measurement error. Due to the sparse nature of the labels (roughly only 4 pixels in 480x640 total pixels per frame), the modified mean squared error loss allows us to independently penalize false positive loss and false negative loss by a hyperparameter that can be chosen based on class imbalance.

$$l_{wmse} = \frac{1}{n}\sum_{i=0}^{n}[\alpha(1-y_i)(y_i-\hat{y}_i)^2 + (1-\alpha)y_i(y_i-\hat{y}_i)^2]$$

An alpha value of 0.001 and an augmented L2 loss weight of 0.001 were chosen through experimentation and hyperparameter sweeps. This allows the model to predict more pixels with high confidence improving stability, training time and interpretability.

The centroid of each output channel is used as the endpoints for each measurement. In beat-to-beat evaluation, heuristics were used to exclude low quality frames of the video from overall calculation. At inference time, low confidence pixels (with a score less than 0.3) were ignored in calculating the centroid. If the model predicts measurements that are inconsistent in angle (greater than 30 degrees different), the frame is considered low quality and the prediction is ignored during video-wide averaging.